%% file: ms.tex
\documentclass[rmp,twocolumn, superscriptaddress]{revtex4-1}

\usepackage{amsmath}
\usepackage{makeidx}
\usepackage{amsfonts}
\usepackage{graphicx}  
\usepackage{bm}        
\usepackage{amssymb}   
\usepackage{color}
\include{commands}

\definecolor{drab}{rgb}{0.59, 0.44, 0.09}
\newcommand{\Richard}[1]{{\color{drab}Richard: #1}}

\begin{document}
\title{Phylodynamics of rapidly adapting pathogens: extinction and speciation of a Red Queen.}
\author{Le Yan\thanks{lyan@kitp.ucsb.edu}}
\affiliation{Kavli Institute for Theoretical Physics, UC Santa Barbara}
\author{Richard A.~Neher}
\affiliation{Biozentrum, University of Basel, Switzerland}
\author{Boris I.~Shraiman}
\affiliation{Kavli Institute for Theoretical Physics, UC Santa Barbara}

\date{\today}

\begin{abstract}
Rapidly evolving pathogens like influenza viruses can persist by accumulating antigenic novelty fast enough to evade the adaptive immunity of the host population, yet without continuous accumulation of genetic diversity. This dynamical state is often compared to the Red Queen evolving as fast as it can just to maintain its foothold in the host population: Accumulation of antigenic novelty is balanced by the build-up of host immunity. Such Red Queen States (RQS) of continuous adaptation in large rapidly mutating populations are well understood in terms of Traveling Wave (TW) theories of population genetics. Here we shall make explicit the mapping of the established Multi-strain Susceptible-Infected-Recovered (SIR) model onto the TW theory and demonstrate that a pathogen can persist in RQS if cross-immunity is long-ranged and its population size is large populations allowing for rapid adaptation. 
We then investigate the stability of this state focusing on the rate of extinction and the rate of ``speciation" defined as antigenic divergence of viral strains beyond the range of cross-inhibition. RQS states are transient, but in a certain range of evolutionary parameters can exist for the time long compared to the typical time to the most recent common ancestor ($T_{MRCA}$). 
In this range the steady TW is unstable and the antigenic advance of the lead strains relative to the typical co-circulating viruses tends to oscillate. This results in large fluctuations in prevalence that facilitate extinction. We shall demonstrate that the rate of TW fission into antigenically uncoupled viral populations is related to fluctuations of  $T_{MRCA}$  and construct a ``phase diagram" identifying different regimes of viral phylodynamics as a function of evolutionary parameters.
\end{abstract}

\maketitle
In a host population that develops long lasting immunity against a pathogen, the pathogen can persist either by infecting immunological naive individuals such as children or through rapid antigenic evolution that enables the pathogen to evade immunity and re-infect individuals.
Childhood diseases like measles or chicken pox fall into the former category, while influenza virus populations adapt rapidly and reinfect most humans multiple times during their lifespan.
Continuous adaptation of influenza viruses is facilitated by high mutation rates and it is common that many different variants of the same subtype co-circulate.
Nevertheless, almost always a single variant eventually outcompetes the others such that diversity within one subtype remains limited \citep{petrova_evolution_2018}.

\begin{figure*}[ht]
	\centering
	\includegraphics[width=0.9\textwidth]{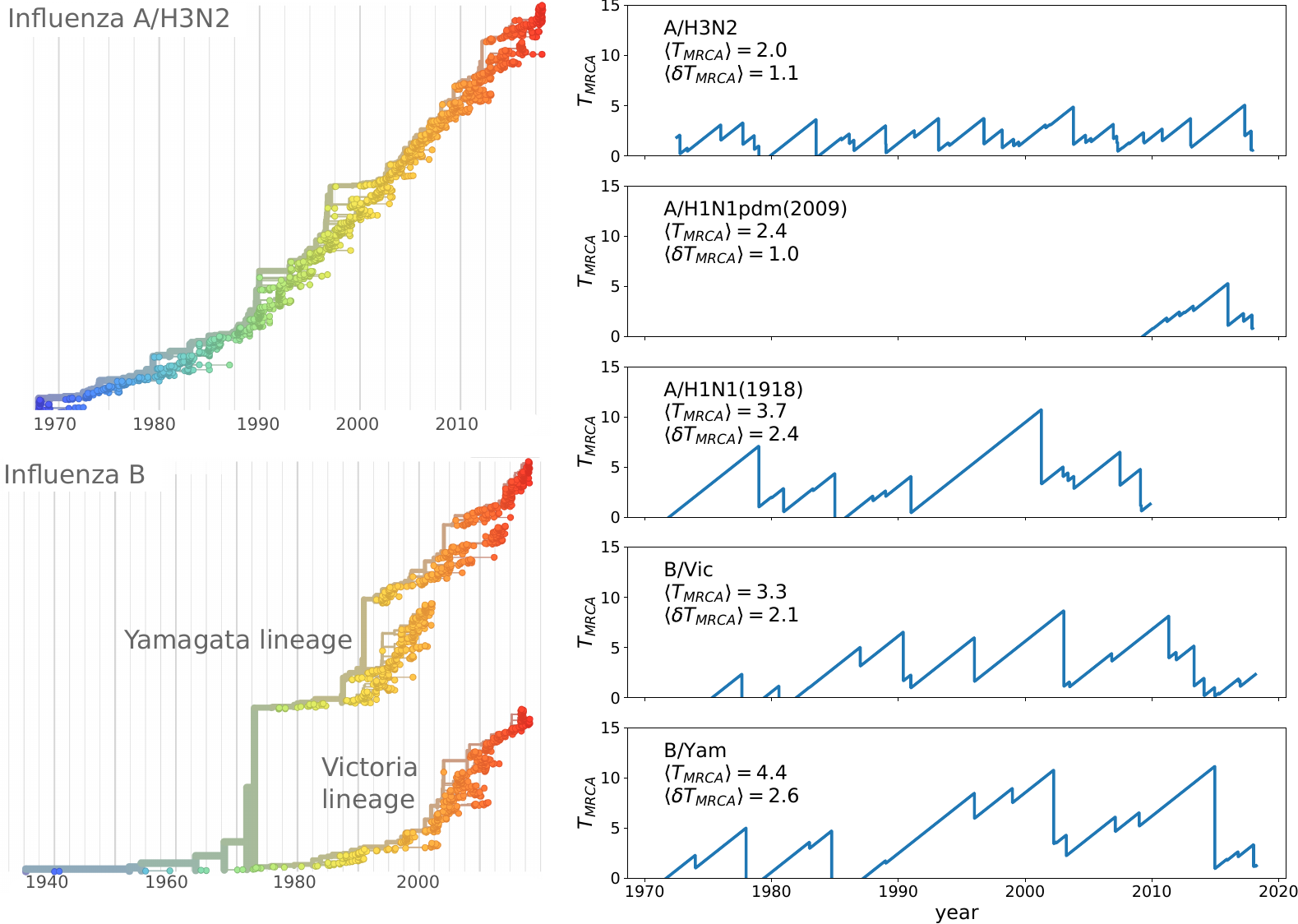}
    \caption{{\bf Spindly phylogenies and speciation in different human seasonal influenza virus lineages.} The top left panel shows a phylogeny of the HA segment of influenza A virus of subtype H3N2 from its emergence in 1968 to 2018. The virus population never accumulates much diversity but is rapidly evolving. The lower left panel shows a phylogeny of the HA segment of influenza B viruses from 1940 to 2018. In the 70ies, the population split into two lineages known as Victoria (B/Vic) and Yamagata (B/Yam). The graphs on the right quantify diversity via the time to the most recent common ancestor $T_{MRCA}$ for different influenza virus lineages. Influenza B viruses harbor more genetic diversity than influenza A viruses. The subtype A/H3N2 in particular coalesces typically in 3y while deeps splits in excess of 5y are rare. }
	\label{fig:influenza}
\end{figure*}

The contrast of rapid evolution while maintaining limited genetic diversity is most pronounced for the influenza virus subtype A/H3N2.
\FIG{influenza} shows a phylogenetic tree of HA sequences of type A/H3N2 with the characteristic ``spindly'' shape.
The most recent common ancestor of the population is rarely more than 3-5 years in the past.
Other pathogenic RNA viruses that typically don't reinfect the same individual, (measles, mumps, HCV or HIV) diversify for decades or centuries \citep{Grenfell04}.
Interestingly, influenza B has split into two co-circulating lineages in the 1970ies which by now are antigenically distinct \citep{rota_cocirculation_1990} and maintain intermediate levels of diversity.
Thus, one may wish to understand under what conditions a virus can continuously evolve in competition with host immunity, neither going extinct nor spawning diverging lineages.

A variety of mechanisms have been proposed that can lead to spindly phylogenies in rapidly evolving viral populations.
A low dimensional antigenic space might limit the number of distinct directions in which the population can evolve thereby preventing diversification \citep{Andreasen97,andreasen_shaping_2006,Gog02}.
This scenario seems inconsistent with estimates of the dimensionality of ``antigenic shape space'' ($d=5$ \citep{perelson_theoretical_1979}) and the number of distinct positions in surface proteins that evolve under immune selection \citep{koel_substitutions_2013,Neher16,bhatt_genomic_2011}.
Others have shown that competition between lineages and long range cross-immunity between strains can prevent diversification, effectively canalizing the population into a single lineage \citep{Bedford12,tria_minimal_2005,Ferguson03}.
Deleterious mutational load also facilitates the maintenance of a single lineage
\citep{Koelle15}.

While these previous studies have identified competition mediated by cross-immunity as important determinants of lineage structure, these prior works mostly relied on simulations with the aim to recapitulate patterns observed for human seasonal influenza viruses.
Here, we show that generic stochastic models of antigenic evolution with finite cross-immunity are compatible with the spindly phylogenies of influenza viruses if the range of cross-immunity is large compared to population diversity.
If cross-immunity decays more rapidly with mutational distance, viral population becomes prone to speciation. 
Finally, if antigenic evolution is too slow, the virus will go extinct after a brief pandemic.
This rich behavior is controlled by three dimensionless parameters that have a direct relationship to the parameters of the multi-strain SIR model.
We show how multi-strain SIR models relate to traveling wave models of adaptive evolution \citep{Tsimring96,rouzine_solitary_2003,Desai07,neher_genetic_2013} and how extinction and speciation relate to oscillations of prevalence and genetic diversity.

\subsection*{Model}
A model of an antigenically evolving pathogen population needs to account for cross-immunity between strains and the evolution of antigenically novel strains.
We use an extension of the standard multi-strain SIR model \citep{Gog02}. The fraction of individuals $I_a$ infected with viral strain $a$ changes according to
\begin{equation}
\label{eq:infected}
\ddt I_a = \beta S_a I_a - (\nu+\gamma)I_a
\end{equation}
where $\beta$ is the transmissibilty, $S_a$ is the fraction of the population that is susceptible to strain $a$, $\nu$ is the recovery rate, and $\gamma$ is the population turnover rate. Susceptibility of strain $S_a$ depends on the fraction $R_b$ of the population recovered from infections with strain $b$
\begin{eqnarray}
\label{eq:susceptible}
S_a  &=&e^{-\sum_b K_{ab} R_b } \\
\label{eq:susceptible2}
\ddt R_a  &=& \nu I_a - \gamma R_a
\end{eqnarray}
The matrix $K_{ab}$ quantifies the cross-immunity to strain $a$ elicited by infection with strain $b$, while Eq.~\ref{eq:susceptible2} describes recovery.
This model approximates the population susceptibility by the average number of infections with each strain $b$ and ignores the explicit infection history of individuals.

\begin{figure}[tb]
	\centering
	\includegraphics[width=0.48\textwidth]{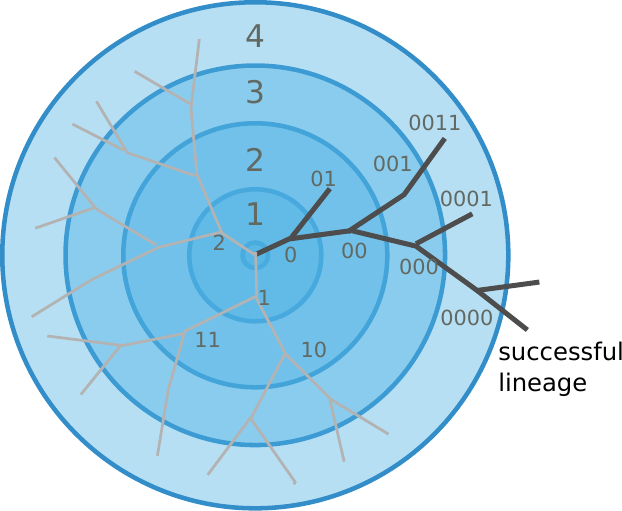}
	\caption{{\bf Viral populations escape adaptive immunity by accumulating antigenic mutations.} Via cross-reactivity, the immunity foot-print of ancestral variants (center of the graph) mediates competition between related emerging viral strains and can drive all but one of the competing lineages extinct. At high mutation rates and relatively short range of antigenic cross-reactivity, different viral lineages can escape inhibition and continue to evolve independently.}
	\label{fig:sketch}
\end{figure}

New strains are constantly produced by mutation with rate $m$.
The novel strain will differ from its parent at one position in its genome.
We shall consider only mutations that contribute to the loss of immune recognition and assume that cross-immunity decays exponentially with the number of mutations that separate two strains:
\begin{equation}
    \label{eq:immunity_kernel}
    K_{ab} = e^{-\frac{|a-b|}{\CI}}
\end{equation}
where $|a-b|$ denotes the mutational distance between the two strains and $\CI$ denotes the radius of cross-immunity measured in units of mutations.
Antigenic space is thereby assumed to be high dimensional and antigenic distance is proportional to genetic distance in the phylogenetic tree \citep{Neher16}.
The distance between two contemporaneous strains is on average twice the distance to their common ancestor.

Cross-immunity and the mutation/diversification process are illustrated in \FIG{sketch}.
An infection with a virus (center of the graph) generates a cross-immunity footprint (shaded circles).
Mutation away from the focal strain reduces the effect of existing immunity in the host population, but complete escape requires many mutations.
Hence closely related viruses compete against each other for susceptible individuals.

The above model was formulated in terms of the deterministic Eqs.~\ref{eq:infected} and \ref{eq:susceptible}.
The actual dynamics, however, is stochastic in two respects:
i) antigenic mutations are generated at random with rate $\mut$
and ii) stochasticity of infection and transmission. 
This stochasticity can be captured by interpreting the terms in Eq.~\ref{eq:infected} as rates of discrete transitions in a total population of $N_h$ hosts.
The latter manifestation of stochasticity is particularly important for novel mutant strains that are rare.
The great majority of novel strains are quickly lost to stochastic extinction even if they have a growth advantage due to antigenic novelty.
To account for stochasticity in a computationally efficient way, we employ a clone-based hybrid scheme where mutation and the dynamics of rare mutants is modeled stochastically, while common strains follow the deterministic dynamics, see Clone-based simulation in Methods.
To simplify the analysis, we will assume that the population turn-over rate $\gamma$ is small compared to other time scales of the system and we will set $\gamma=0$.
We will use the recovery rate $\nu$ to set the unit of time, fixing $\nu=1$ in rescaled units.
The remaining parameters of the model are 1) the transmission rate $\beta$ - in our units the number of transmission events per infection and hence equal to the basic reproduction number $R_0$, 2) the mutation rate $\mut$, 3) the range of cross-immunity $d$ measured as a typical number of antigenic mutations needed for an $e$-fold drop of cross-inhibition, and 4) the host population size $N_h$, which controls the number of opportunities for adaptive evolution and the time it takes for a newly mutated strain to reach population frequency (i.e. prevalence) of order one.

\begin{figure*}
    \centering
    \includegraphics[width=1.\columnwidth]{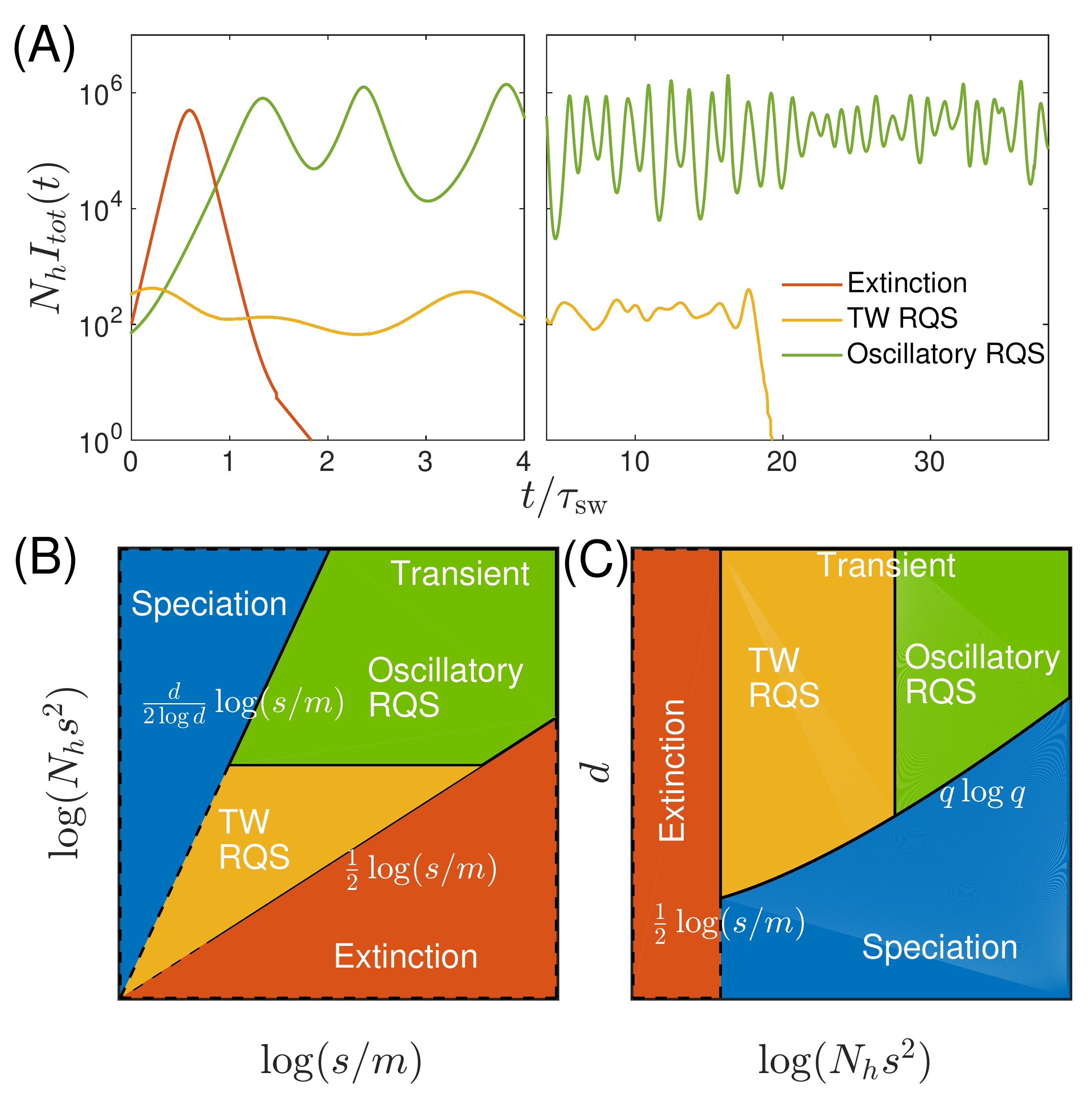}
    \includegraphics[width=.82\columnwidth]{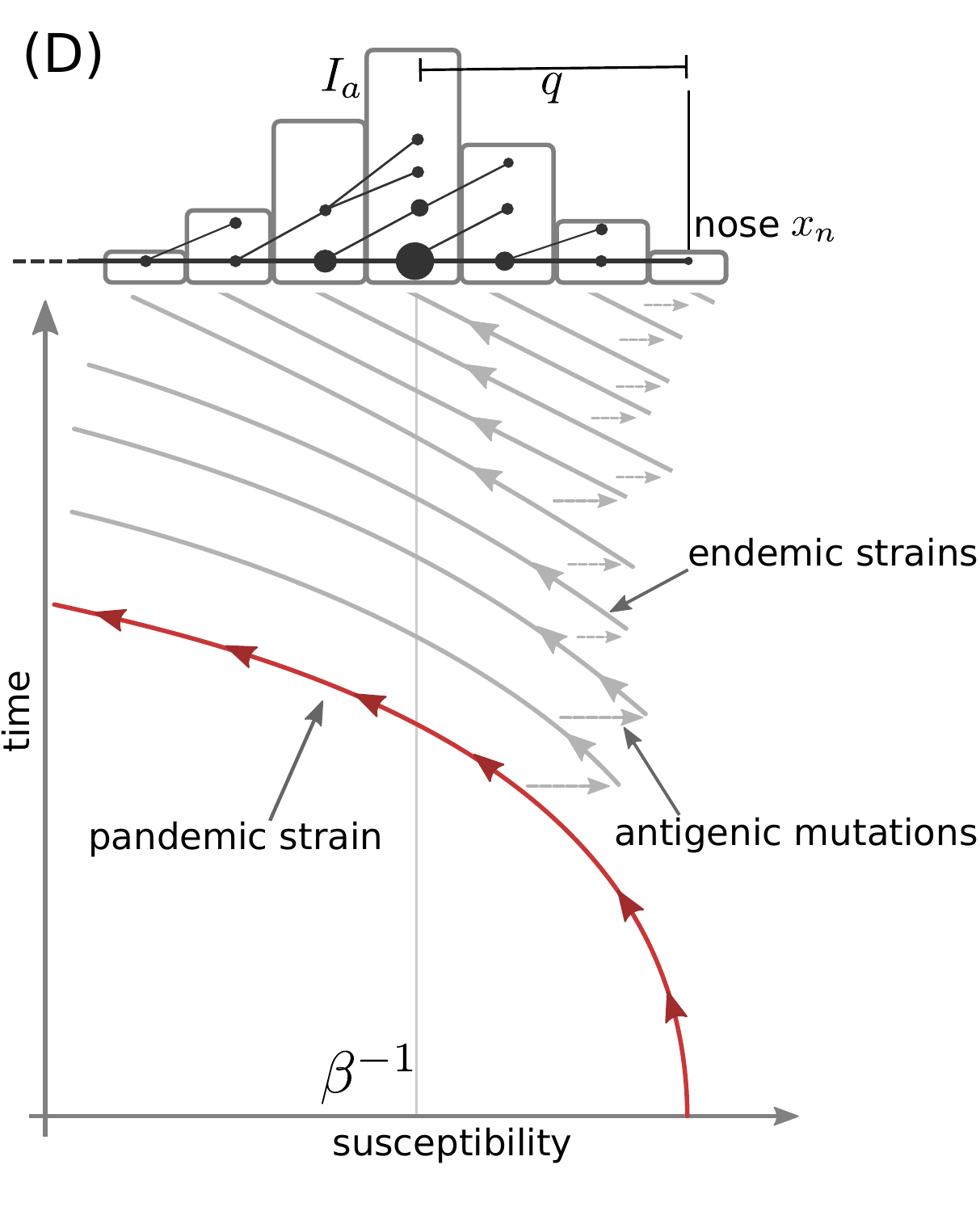}
\caption{{\bf Phases in the multi-strain SIR simulation and the transition from the pandemic to endemic.} (A) Typical trajectories of infection prevalence in the regime of extinction (red),  traveling wave RQS (yellow) and oscillatory RQS (green) which are realized for different values of model parameters ($N_h$ and $m$).
(B,C) The extended multi-strain SIR model supports a long-lived but transient RQS regime in an intermediate range of parameters, flanked be the regime of deterministic extinction (red) and the regime of continuous branching and diversification -- the ``speciation" regime (blue). Panel (B) shows dependence on population size and mutation rate: small populations at low mutation rates go extinct, while large and rapidly mutating populations continuously produce divergent lineages (speciation). Simulation results supporting this phase diagram are shown in \ref{fig:phasedata}. 
Panel (C) presents dependence on the range of antigenic cross-inhibition ($d$) and the population size. The range of cross inhibition controls the crossover between RQS and speciation regimes. The RQS regime itself undergoes a transition from a steady travelling wave (yellow) to a limit cycle oscillation (green) with increasing population size. 
Panel (D) is a schematic illustrating the dynamics of the RQS state. A novel pandemic strain (red) initially expands in fully susceptible population. As the cumulative number of infected individuals increases, the susceptible fraction decreases, and survival of the strain depends on the emergence of antigenic escape mutations (grey). The pathogen may then settle into an endemic, Red Queen-type state where new antigenic variants are continuously produced. The top part of the panel illustrates the population composition at a particular time point. Rare pioneering variants are $q$ mutations ahead of the dominant variant and grow with rate $x_n$. Different lineages are related via their phylogenetic tree embedded in the fitness distribution in the population.
 }
    \label{fig:phase}
\end{figure*}

Before proceeding with a quantitative analysis we discuss different behaviors qualitatively.
\FIG{phase}A shows several trajectories of prevalence  $\Itot = \sum_a I_a$ (i.e.~total actively infected fraction) for several different parameters.
Depending on the range of cross-immunity, the pathogen either goes extinct after a single pandemic (red lines) or settles into a persistently evolving state, the Red Queen State (RQS) traveling wave~\cite{vanValen1973}. 
In large populations the RQS exhibits oscillation in prevalence. The RQS is only transient, but its lifetime increases with the host population size $N_h$ and the mutation rate $\mut$. 
To quantitatively understand the dependence on parameters (shown schematically in \FIG{phase}BC), we will further simplify the model and establish a connection to models of rapid adaptation in population genetics.


\subsection*{Large effect antigenic mutations allow transition from pandemic to seasonal dynamics}
A novel virus in a completely susceptible population will initially spread with rate $\beta-1$ and the pandemic peaks when susceptible fraction falls to $ \beta^{-1}$.
The trajectory of such an pandemic strain in the time-susceptibility plane is indicated in red in \FIG{phase}D.
Further infections in the contracting epidemic will then push susceptibility below $\beta^{-1}$ -- the propagation threshold for the virus -- and without rapid antigenic evolution the pathogen will go extinct after a time $t\sim \beta^{-1}\log N_h$.
Such boom-bust epidemics are reminiscent of the recent Zika virus outbreak in French Polynesia and the Americas where in a short time a large fraction of the population was infected and developed protective immunity \citep{oreilly_projecting_2018}.

Persistence and transition to an endemic state is only possible if the pathogen can evade the rapid build-up of immunity via a small number of large effect antigenic mutations.
This process is indicated in \FIG{phase}D by horizontal arrows leading to antigenically evolved strains of higher susceptibility and bears similarity to the concept of ``evolutionary rescue'' in population genetics \citep{gomulkiewicz_when_1995}.
The parameter range of the idealized SIR model that  avoid extinction after a pandemic resulting in persistent endemic disease is relatively small.
Yet, various factors like geographic structure, heterogeneity of host adaptation and population turn-over slow down the pandemic and extinction, thereby increasing the chances of sufficient antigenic evolution to enter the endemic, RQS-type, regime.
The 2009 pandemic influenza A/H1N1 has undergone such a transition from a pandemic to a seasonal/endemic state.
We shall not investigate the transition process in detail here, but will assume that endemic regime has been reached.

\subsection*{Long range cross-immunity results in evolving but low diversity pathogen populations}
Once the pathogen population has established an endemic circulation through continuous antigenic evolution (green and yellow regimes in \FIG{phase}), the average rate of new infections $\beta \sum_a I_a S_a/\Itot$ fluctuates around the rate of recovery $\nu = 1$ (in our time units).
This balance is maintained by the steady decrease in susceptibility due to rising immunity against resident strains and the emergence of antigenically novel strains, see \FIG{phase}D.
If the typical mutational distance between strains is small compared to the cross-immunity range $\CI$, the rate at which susceptibility decreases is similar for all strains. To see this we differentiate Eq.~\ref{eq:susceptible} with respect to $t$ and divide by $S_a$:
\begin{equation}
	S_a^{-1}\ddt S_a(t) = - \sum_b e^{-\frac{|a-b|}{d}} I_b \approx - \Itot +\sum_b \frac{|a-b|}{d}I_b
\end{equation}
where we have used that $|a-b|\ll d$ for all pairs of strains with substantial prevalence. In fact it will suffice to keep only the first, leading, term on the right hand side.
Close to a steady state, prevalent strains obey $\beta S_a \approx 1$.
We can hence define the instantaneous growth rate of strain $x_a=(\beta S_a - 1) \ll 1$ as its effective fitness.
In this limit, the model can be simplified to
\begin{equation}
    \label{eq:TW}
    \begin{split}
        \ddt I_a  &= x_a I_a \\
        \ddt x_a  &\approx  - \Itot
    \end{split}
\end{equation}
The second equation means that effective fitness  of all strains $a$ decreases approximately at the same rate since the pathogen population is dominated by antigenically similar strains.

If a new strain $c$ emerged from strain $a$ by a single antigenic mutation, its mutational distance from a strain $b$ is $|c-b| = |a-b|+1$ and $K_{cb}=K_{ab}e^{-d^{-1}}\approx K_{ab}(1-d^{-1})$.
The population susceptibility of strain $c$ is therefore increased to
\begin{equation}
    S_c \approx e^{-(1-d^{-1})\sum_b K_{ab}R_b}
    \approx S_a\left(1-\frac{\log S_a}{d}\right)
\end{equation}
Since the typical susceptibility is of order $\beta^{-1}$, the growth rate of strain of the mutant $c$ is $s = d^{-1}\log \beta$ higher than that of its parent.
The growth rate increment plays the role of a selection coefficient in typical population genetic models and corresponds to the step size of the fitness distribution in \FIG{phase}D.
In this model, individuals within a fitness class (bin of the histogram) are equivalent and different classes can be modeled as homogeneous populations which greatly accelerates numerical analysis of the model, see Methods. 

The simplified model in Eq.~\ref{eq:TW} is analogous to the traveling wave (TW) models of rapidly adapting asexual populations that have been studied extensively over the past two decades \citep{Tsimring96,Desai07,rouzine_solitary_2003,hallatschek_noisy_2011}, see \citep{neher_genetic_2013} for a review.
These models describe large populations that generate beneficial mutations rapidly enough that many strains co-circulate and compete against each other.
The fittest (most antigenically advanced) strains are often multiple mutational steps ahead of the most common strains.
This ``nose'' of the fitness distributions contains the strains that dominate in the future and the only adaptive mutations that fixate in the population arise in pioneer strains in the nose.
Consequently, the rate with which antigenic mutations establish in the population is controlled by the rate at which they arise in the nose \citep{Desai07}.
If the growth rate at the nose of the distribution, $x_n$, is much faster than antigenic mutation rate, $x_n \gg m$ it takes typically
\begin{equation}
	\tau_a = \frac{\log (x_n/\mut)}{x_n}
\end{equation}
generations before a novel antigenic mutation arises in a newly arisen pioneer strain that grows exponentially with rate $x_n $.
The advancement of the nose is balanced rapidly by the increasing population mean fitness.
Previous analyses have shown that in the limit of very large population $N\gg1$, the fitness distribution has an approximately Gaussian shape with a variance $\sigma^2 \approx 2s^2{\log (N s)}/ \log^2 (x_n/m)$.
The wave is $\sigma /s$ mutations wide, while the most advanced strains are approximately $q = 2 \log (N s)/\log (x_n/m)$ ahead of the mean \citep{Desai07}.
Two contemporaneous lineages coalesce on a time scale $\tau_{\rm sw} = sq/\sigma^2=s^{-1} \log (x_n/m)$ and the branching patterns of the tree resemble a Bolthausen-Sznitman coalescent rather than a Kingman coalescent \citep{desai_genetic_2013,neher_genealogies_2013}. 
We note that, while parameter $N$ in the TW analysis summarized above is the fixed population size, the corresponding entity in our SIR model is the pathogen population size $N_p$, which is related to the (fixed) host population size $N_h$ by $N_p=N_h {\bar I}$ where $\bar I$ is the average viral prevalence, which itself depends on other parameters of the model, scaling in particular with $s^2$. Hence, it will be convenient for us to use $N_hs^2$ as one of the relevant ``control parameters", replacing $N$ of the standard TW model.
A recent related work, that also explicitly maps a multi-strain SIR models to the TW models, but does not consider the role of population size fluctuations \citep{rouzine_antigenic_2018}.

\begin{figure}
    \centering
    \includegraphics[width=\columnwidth]{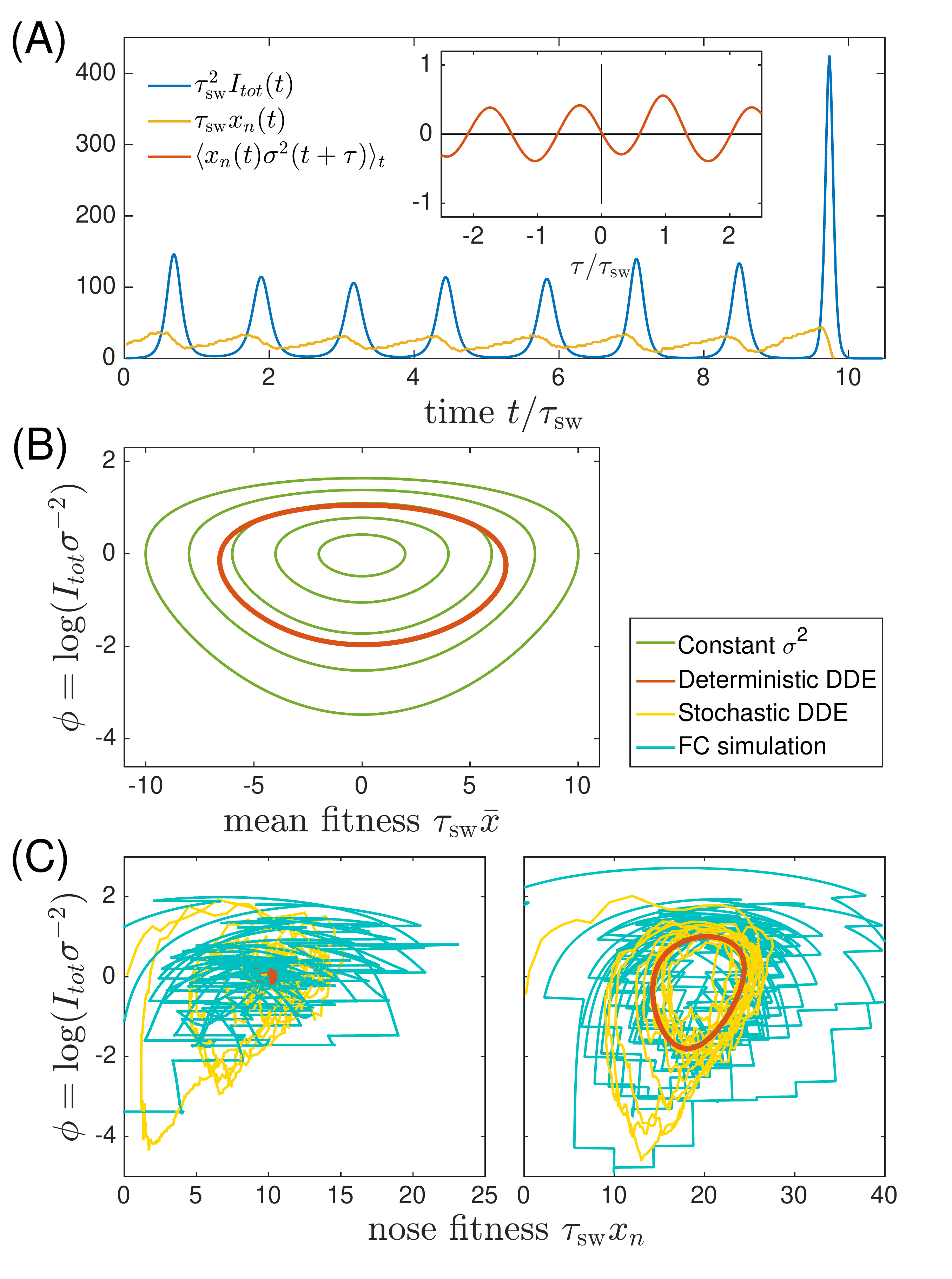}
    \caption{{\bf Oscillatory RQS.} (A)  An example of the stochastic limit cycle trajectory from the fitness-class simulation. Note the rapid rise and fall of infection prevalence (blue), which causes a drop in nose fitness (yellow) which subsequently recovers (approximately linearly) during the remainder of the cycle. Fluctuations in $I_{tot}(t)$ and $x_n(t)$ from cycle to cycle are caused by the stochasticity of $x_n$, i.e.~antigenic evolution in pioneer strains. A particularly large fluctuation about $\tau_{\rm sw}$ prior to the end, caused a large spike in prevalence, followed by the collapse of $x_n$ below zero and complete extinction. Inset (red) shows the cross-correlation between $x_n$ and $\sigma^2$ which peaks with the delay $\tau=\tau_{\rm sw}$ (additional peaks reflect the oscillatory nature of the state and are displaced by integer multiples of mean period); (B) A family of limit cycles in infection prevalence/mean fitness plane as described by \EQ{mean} with fixed variance. The variation of $\sigma$ governed by the Eqs.~\ref{eq:mean}-\ref{eq:nose} (in the deterministic limit) reduces the family to a single limit cycle (red); (C) Trajectories in the infection prevalence/nose fitness generated by the stochastic DD system in the regime above (right panel) and below (left panel) the oscillatory instability of the deterministic DD system.}
    \label{fig:I_vs_nose}
\end{figure}

\subsection*{Stability and fluctuations of the RQS}
In contrast to most population genetic models of rapid adaptation, our epidemiological model  does not control the total population size directly. 
Instead, the pathogen population size (or prevalence) depends on the host susceptibility, which in itself is determined by recent antigenic evolution of the pathogen.
The coupling of these two different effects results in a rich and complicated dynamics:
The first effect is ecological: a bloom of pathogen depletes susceptible hosts leading to a crash in pathogen population and a tendency of the population size to oscillate~\cite{London1973}. 
The second effect is evolutionary: higher nose fitness $x_n$ begets faster antigenic evolution and vice versa, resulting in an apparent instability in the advancement of the antigenic pioneer strains.
This instability has been identified in the study of adaptive traveling waves \citep{Fisher13}.
In our epidemiological model, fluctuations in the rate of antigenic advance of the pioneer strains couple, with a delay of $\tau_{\rm sw}$, to the ecological oscillation.

To recognize the origin of the oscillatory tendency, consider the total prevalence $\Itot$ and the mean fitness of the pathogen $X=\sum_a x_a I_a /\Itot$
\begin{equation}
\frac{d}{dt}  \Itot=X\Itot; \ \ \ \ \ \frac{d}{dt} X = \sigma^2 - \Itot
\label{eq:mean}
\end{equation}
At fixed variance $\sigma={\bar \sigma}$ this system is equivalent to a non-linear oscillator, describing a family of limit cycles oscillating about $\Itot ={\bar \sigma}^2 $ and $X=0$ as shown in \FIG{I_vs_nose}B.

While \EQ{mean} describes the behavior of common strains, the dynamics of the antigenic pioneer strains is governed by the equation for $x_n$ that in a continuum limit (suitable for the limit of high mutation rate) reads: 
\begin{equation}
\frac{d}{dt} x_n = \tau_{\rm sw}^{-1}x_n  - \Itot + s \xi (t)
\label{eq:nose}
\end{equation}
The first term on the right hand side represents the rate with which antigenic pioneer strains enter the population, $\tau_a^{-1}$, advancing the nose fitness by an increment $s$ ($\tau_a^{-1}s = \tau_{\rm sw}^{-1}x_n$).
The second term on the right hand side of \EQ{nose} represents gradual reduction of susceptibility of the host population, and $\xi(t)$ is a random noise variable representing the stochasticity of the establishment of new strains.
(The Gaussian white noise $\xi (t)$ is defined statistically by its correlation function $\langle\xi(t) \xi(0) \rangle = \tau_a^{-1} \delta (t)$, see Methods. )

The first term of \EQ{nose} captures the apparent instability of the nose:  an advance of the nose to higher $x_n$  accelerates its rate of advancement. The stabilizing factor is the subsequent increase in $\Itot$, but to see how that comes about we must connect \EQ{nose} to \EQ{mean}. The connection is provided by $\sigma^2$
since it is controlled by the emergence of novel strains, i.e.~the dynamics of the ``nose'' $x_n$, which impacts the bulk of the distribution after a delay $\tau_{\rm sw}$.
Based on the analysis detailed in the Appendix A we approximate
\begin{equation}
\sigma^2 (t) \approx \tau_{\rm sw}^{-1}x_n(t-\tau_{\rm sw})
\label{eq:sigma_delay}
\end{equation}
relating population dynamics, \EQ{mean}, to antigenic evolution of pioneer strains described by \EQ{nose}. Taken together Eqs.~(\ref{eq:mean}-\ref{eq:sigma_delay}) define a Differential Delay (DD) system of equations derived in Appendix A. Sample simulations of this stochastic DD system are shown in \FIG{I_vs_nose}(BC). The delay approximation \EQ{sigma_delay} is supported by the cross-correlation of $x_n(t)$ and $\sigma^2(t')$ measured using fitness-class simulations (see \FIG{I_vs_nose}A Inset)

\begin{figure*}
    \centering
    \includegraphics[width=1.3\columnwidth]{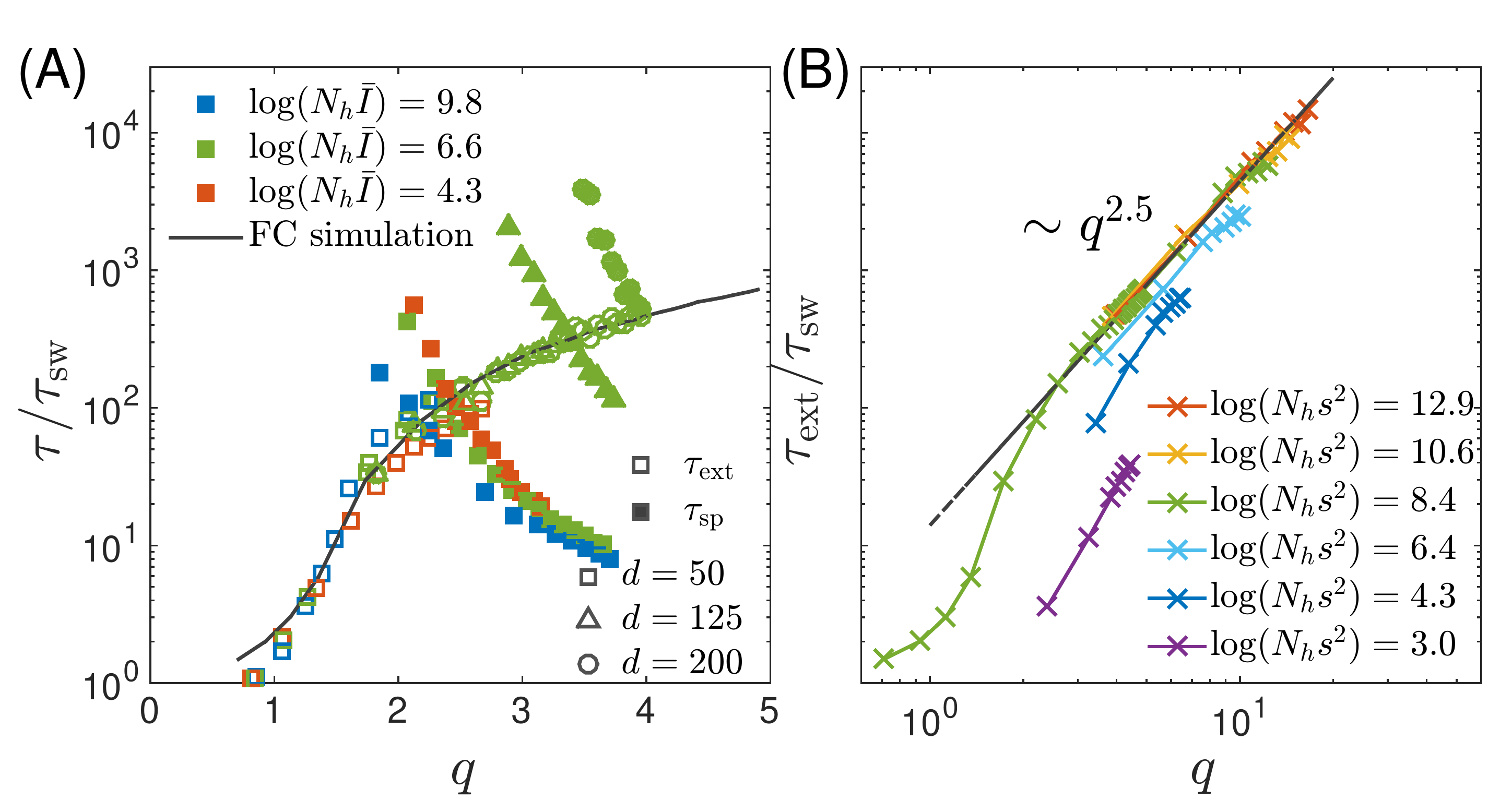}
\caption{(A) Simulation results for the average extinction time and the average speciation time obtained in the clone-based simulation. Extinction time $\tau_{\rm ext}$, scaled with the sweep time $\tau_{\rm sw}$, increases with the depth of the genealogy measured by the number of mutations $q$ separating the most antigenically advanced strain from the most common strain. Also shown is the characteristic time to speciation $\tau_{\rm sp}$, which increases with the range of cross-inhibition and decreases with $q$. The crossover of the two time scales defines the transition from transient RQS to speciation. (B) Extinction time over a broad range of parameters, obtained via fitness class-based simulation of population dynamics, confirms its primary dependence on $q$ for large population sizes. Note the agreement between the results of the fitness class-based simulation (black line in (A)) and the clone-based simulation (colored squares in (A)). }
    \label{fig:extinction}
\end{figure*}

The deterministic limit of the DD system (obtained by omitting the noise term in \EQ{nose}) has two qualitatively different regimes that correspond to the TW and oscillatory regimes. 
Small deviations from the steady state with $\tau_{\rm sw}^{-1}\bar x_n=\bar\sigma^2=2\tau_{\rm sw}^{-2}\log (N_h\bar{I})$, are underdamped and oscillate with frequency  $\omega={\bar\sigma }={\tau_{\rm sw}}^{-1}{\sqrt{2\log (N_h\bar{I})}}$ (as determined by linearizing \EQ{mean}). When  $\omega \tau_{\rm sw} < 2\pi $, the delayed feedback via $\Itot$ stabilizes the steady state, while in the opposite regime, the system fails to recover from a deviation of the nose in a single period and the steady state becomes unstable to a limit cycle oscillation. As the nonlinearity of \EQ{mean} implies a longer period with increasing amplitude, the system is stabilized at a limit cycle with the period long enough compared to the feedback delay $\tau_{\rm sw}$. In Appendix A we derive the threshold of oscillatory instability to lie at  $\log (N_h \bar{I}_{osc}s) \approx 8.3$ (leading to limit cycle period $T \approx 1.5\tau_{\rm sw}$, see \FIG{variance} in SI). We also find that the amplitude of the oscillation $\log(I_{max}/{\bar I})$ scales as $\log (N_h\bar{I})$ for large values of the later.
This transition defines quantitatively the boundary between the TW RQS and the Oscillatory RQS regimes that appear on the phase diagrams in \FIG{phase}(BC).

The distinction between the TW and Oscillatory RQS is obscured by the stochasticity of antigenic advance, \EQ{nose}, which continuously feeds the underdamped relaxation mode, generating a noisy oscillation with the frequency $\omega$ defined above. The difference between the two regimes is illustrated by \FIG{I_vs_nose}C: in the TW RQS noisy oscillation is about the fixed point, whereas in the Oscillatory RQS it is about deterministic limit cycle.

Interestingly, the dynamics of the Oscillatory RQS, as shown in \FIG{I_vs_nose}A, can be understood in terms of a non-linear relaxation oscillator. At relatively low infection prevalence nose fitness $x_n$ increases until rising $\Itot$ catches up with it (when $\Itot=\tau_{\rm sw}^{-1}x_n$) driving it down rapidly. Once this ``mini-pandemic" burns out, the population returns to the low prevalence part of the cycle $\Itot<\tau_{\rm sw}^{-1}x_n$, when $x_n$ begins to increase again. This relaxation oscillator approximation is discussed in more detail in the Supplementary Information.

\subsection*{The rate of extinction}

While in the deterministic limit the differential-delay system predicts a stable steady TW (for $q>q_{ex}, \ \bar{I}<\bar{I}_{osc}$) and a limit cycle (above $\bar{I}_{osc}$), (see SI) fluctuations in the establishment of the antigenic pioneer strains (\EQ{nose}) can lead to stochastic extinction. In fact, both the TW and Oscillatory RQS (see Fig. 3BC) are transient, subject to extinction due to a sufficiently large stochastic fluctuation. (Note however the contrast with the ``Extinction" state in \FIG{phase}BC, where extinction is deterministic and rapid.)
The rate of extinction depends on $q$ and $\log (N_h\bar{I})$ as shown in Figure \ref{fig:extinction}A, which compares results of the clone-based simulations of the multi-strain SIR model with the fitness class-based simulation.
Although extinction is fluctuation driven, the mechanism of extinction in the oscillatory state is related closely to the deterministic dynamics, according to which large amplitude excursion in infection prevalence  can lead to extinction.  A large $x_n$ advance leads, after a time $\tau_{\rm sw}$ to a rise in prevalence $\Itot$, followed by the rapid fall in the number of susceptible hosts and hence loss of viral fitness. This turns out to be the main mode of fluctuation driven extinction as illustrated by \FIG{I_vs_nose}C.
One expects extinction to take place when the fluctuation induced deviation of $x_n$ (from its mean)   $\delta x \approx s \sqrt{\tau_{\rm ext}/\tau_a}$ becomes of the order of the mean $\delta x \approx {\bar x}_n \theta (\log (N_h \bar{I})) $ with  the extinction threshold $\theta (\log(N_h\bar{I})) $ dependent on the shape of the oscillatory limit cycle (as it depends on the minimum of infection prevalence  during the cycle). This argument suggests $\tau_{\rm ext}/\tau_{\rm sw} \sim f(\sqrt{q} \theta (\log(N_h \bar{I})))$ -- a functional relation borne out by the results of numerical simulations in \FIG{extinction}. We note that the rate increase in $\tau_{\rm ext}$ with increasing $q$ slows down in the oscillatory regime and appears to approach a power law dependence $\tau_{\rm ext}/\tau_{\rm sw} \sim q^{2.5}$ (albeit over a limited accessible range): presently we do not have analytic understanding of this specific functional form.

\begin{figure*}
    \centering
    \includegraphics[width=1.\textwidth]{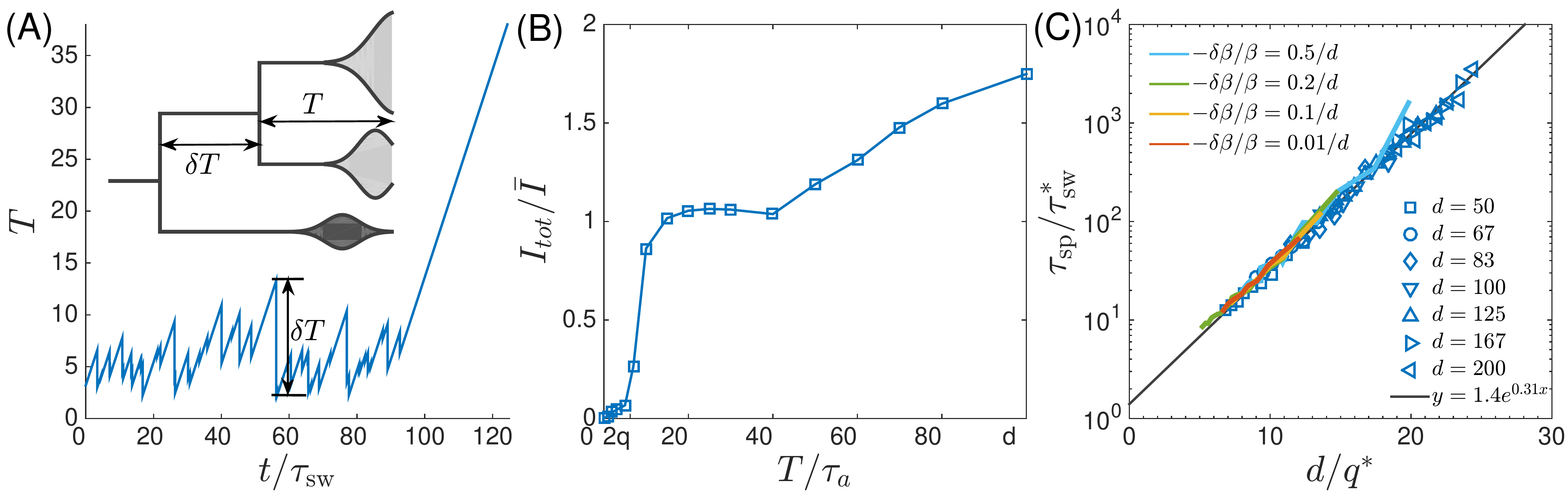}
    \caption{{\bf Speciation into antigenically distinct lineages}. (A) To speciate, two lineage have to diverge enough to substantially reduce cross-reactivity, i.e., $T$ needs to be comparable to $d$. Inset: Illustration of the definition of time to most recent common ancestor $T$ and the time interval $\delta T$ by which $T$ advances. (B) If such speciation happens, the host capacity - the average number of infected individuals increases two-fold. (C) The probability of such deep divergences decreases exponentially with the ratio $d/q^*$, where effective antigenic diversity is $q^*=2\log(N_hs^2)/\log(s/m)$.
    In the presence of deleterious mutations, the relevant $q$ is not necessarily the total advance of the pioneer strains, but only the antigenic contribution. This antigenic advance $q^*$ can be computed as $q^*=\sqrt{2\log(N_hs^2)\sigma_{ag}^2}$ with antigenic variance $\sigma_{ag}^2=\sigma^2-\sigma^2_{\beta}$, where $\sigma^2_\beta$ is fitness variance due to deleterious mutations. With this correction,  speciation times agree with the predicted dependence (colored lines). }
    \label{fig:speciation}
\end{figure*}


\subsection*{The rate of speciation}
The correspondence of the multi-strain SIR and the TW models discussed above assumes that cross-immunity decays slowly compared to the coalescent time of the populations, i.e., $d/q\gg 1$.
In this case, all members of the population compete against each other for the same susceptible hosts. Conversely, if the viral population were to split into two sub-populations separated by antigenic distance greater than the range of cross-inhibition $d$, these sub-population would no-longer compete for the hosts, becoming effectively distinct viral ``species'' that propagate (or fail) independently of each other.
Such a split has for example occurred among influenza  B viruses, see \FIG{influenza}.

A ``speciation" event corresponds to a deep split in the viral phylogeny, with the $T_{MRCA}$ growing without bounds, see \FIG{influenza} and \FIG{speciation}A.
This situation contrasts the phylogeny of the single competing population, where $T_{MRCA}$ fluctuates with a characteristic ramp-like structure generated by stochastic extinction of one of the two oldest clades.
In each such extinction event the MRCA jumps forward by $\delta T$.
Hence the probability of speciation depends on the probability of the two oldest clades to persist without extinction for a time long enough to accumulate antigenic divergence in excess of $d$. 
The combined carrying capacity of the resulting independent lineages is then twice their original carrying capacity as observed in simulations, see \FIG{speciation}B.

To gain better intuition into this process let's follow two most antigenically advanced ``pioneer strains''. In the TW approximation one of these will with high probability belong to the backbone giving the rise to the persisting clade, while the other clade will become extinct, unless it persist long enough to diverge antigenically beyond $d$, becoming a speciation event.
As their antigenic distance gradually increases, the two clades are evolving to evade immunity built up against the common ancestor. The less advanced of the two clades is growing less rapidly and takes longer to  generate antigenic advance mutations, resulting in still slower growth and slower antigenic advance.
Deep splits are hence unstable and it is rare for a split to persist long enough for speciation.
In Appendix B we reformulate this intuition mathematically as a ``first passage''-type problem which shows that $T_{MRCA}$ distribution has an exponential tail which governs the probability of speciation events.
The propensity to speciate depends on the radius of cross-immunity $\CI$ and the typically genetic diversity $q$.
\FIG{speciation} shows that the time to speciation increases approximately exponentially with the ratio $d/q$. More precisely we found that average simulated speciation time  behaves as $ \tau_{\rm sw}^* e^{f(d/q^*)}$ with ``effective'' $\tau_{\rm sw}^*=\tau_{\rm sw}/(1+\log q /\log(s/m))$ and $q^*=q(1+\log q /\log(s/m))$ picking up an additional logarithmic dependence on parameters, the exact origin of which is beyond our current approximations. This correction plausibly suggests rapid speciation, $\tau_{\rm sw}^* \rightarrow 0$, when mutation rate become comparable to the selection strength $m/s \rightarrow 1$.

We emphasize that the line in the ``phase diagram" (\FIG{phase}BC) separating the RQS domain from the Speciation domain refers to the crossover between the regime where RQS is more likely to go extinct before speciating and vice versa. \FIG{extinction}A shows average speciation and extinction times for different parameters -- the intersection of the two times defines the boundary in the phase diagram.

\section*{Discussion}
The epidemiological and evolutionary dynamics of human RNA viruses show a number of qualitatively distinct patterns.
While agents of classical childhood diseases like measles or mumps virus show little antigenic evolution, others viruses like dengue- or norovirus exist in distinct serotypes, while seasonal influenza viruses undergo continuous antigenic evolution enabling the viruses of the same lineage to reinfect the same individual.

Here, we have integrated classical multi-strain SIR models with stochastic models of adaptation to understand the interplay between the epidemiological dynamics and the accumulation of antigenic novelty.
The former is dominated by the most prevalent strains, while the latter depends critically on rare pioneer strains that become dominant at later times.
The integration of these two different crucial aspects of the epi-evolutionary dynamics allowed to define a ``phase'' diagram that summarizes qualitatively different behavior as a function of the relevant parameter combinations, see \FIG{phase}B\&C.

The phase diagram shows different combinations of key parameters that lead to three distinct outcomes: (1) extinction (red), (2) an evolving but low diversity pathogen population (yellow and green), (3) a deeply branching and continuously diversifying pathogen population (blue).
The key parameters are the size of the population $\log(N_hs^2)$, the ratio of mutational effects and mutation rate $\log(s/m)$, and the cross-immunity range $\CI$.
Large $\CI$ prevents speciation, while rapid mutation and large population sizes facilitate speciation.
Of the different regimes, only extinction (1) and speciation (3) are truly asymptotic.
The intermediate regimes of continuously evolving low diversity pathogen population - the Red Queen State (RQS) - are strictly speaking metastable states which eventually either go extinct or undergo branching, but in a certain regime of parameters can be very long lived.

Outbreaks of emerging viruses that quickly infect a large fraction of the population, as for example the recent Zika virus outbreak in the Americas, fall into regime (1): In 2-3 years, large fractions of the population were infected and have developed long-lasting immunity.
As far as we know, the viral population didn't evolve antigenically to escape this build up of herd immunity and the virus population is not expected to continue to circulate in the Americas \citep{oreilly_projecting_2018}.

Different influenza virus lineages, in contrast, persist in the human population, suggesting that they correspond to parameters that fall into the RQS region of the phase diagram.
Furthermore, the different subtypes display quantitatively different circulation and diversity patterns that allow for a direct, albeit limited, comparison to theoretical models.
We know of four seasonal influenza A lineages: subtype A/H1N1 circulated with interruption from 1918 to 2009, A/H2N2 circulated for about 10 years until 1968, A/H3N2 emerged in 1968 and is still circulating today, and the triple reassortant 2009 H1N1 lineage, called A/H1N1pdm, settled into a seasonal pattern following the pandemic in 2009.
Influenza B viruses have split into two separate lineages (B/Victoria and B/Yamagata) around 1983 \citep{rota_cocirculation_1990}.
Phylogenetic trees of A/H3N2 and the influenza B lineages are shown in \FIG{influenza}.
In addition, the figure shows diversity of five lineages as measured by the instantaneous $T_{MRCA}$ through time.

The influenza B lineages tend to be more genetically diverse than the influenza A lineages with a typical time to the most recent common ancestor of around 6 compared to 3 years, see \FIG{influenza}.
A/H3N2 tends to have the lowest diversity and most rapid population turnover.
This difference in diversity is consistent with influenza B lineages being more prone to speciation.

The typical diversity of these viruses needs to be compared to their rate of antigenic evolution.
Hemagglutination inhibition titers drop by about 0.7-1 log2 per year in A/H3N2 compared to 0.1-0.4 log2 per year for influenza B lineages \citep{Smith04,Bedford14,Neher16}.
Hence the ratio of the time required to loose immunity and $\Tmrca$ is similar for the different lineages, suggesting that the distinct rates of genetic and antigenic evolution can not be used as a straight forward rationalization of the speciation event of Influenza B and the lack of speciation of influenza A lineages.
Nor should such an explanation be expected as there is only a single observation of speciation.

From the phase diagram, we found the most relevant parameters that determine the fate of a pathogen are the antigenic diversity $q$ and the range of cross-immunity $d$.
A previous study by \citet{Koelle15} has implicated deleterious mutation load as a cause of spindly phylogenies.
Our model can readily incorporate the effect of deleterious mutations affecting transmission coefficient $\beta$ by $\delta\beta$ and subsequent compensatory mutations that increase $\beta$.
Such a modification is expected to reduce the average $\beta$ and reduce the selection coefficient of antigenic mutations, which in turn reduces the fitness variance $\sigma^2$, as derived in Supplementary Information.
After subtracting the contribution of deleterious mutations from the the fitness variance, the times to speciation follow the predicted dependence on $q$ and $\CI$, see \FIG{speciation}C.

While we have shown that the natural tendency of SIR models to oscillate  couples to the instability of the nose of the pathogen fitness distribution, making a quantitative link to the observed epidemiological dynamics of the flu is difficult on account of seasonal oscillation in transmissivity. The latter confounding factor is widely believed to be the cause behind observed seasonality of the flu. Including explicit temporal variation (in $\beta$) in our model would lock the frequency of the prevalence oscillation to the seasonal cycle, possibly resulting in subharmonic modulation, yet distinguishing such a modulation on top of an already stochastic process is hard. 
Much remains to be done: finite birth rates, distinct age distributions (as for example is the case for the two influenza B lineages), realistic distribution of antigenic effect sizes, or very long range T-cell mediated immunity would all be interesting avenues for future work.


\begin{table*}[htb]
    \centering
    \begin{tabular}{c|c|c|c}
        Symbol & meaning & comment/range & numerical range \\\hline
        $I_a$ & number of individuals infected with strain $a$ & & \\
        $S_a$ & weighted fraction of individuals susceptible to strain $a$ & $\sim 0.7$ & \\
        $\CI$ & cross-immunity range & $\sim 10$y & $[50,200]$\\
        $\beta=\nu R_0$ & transmission rate & $\sim 3$ & 2\\
        $\nu$ & recovery rate & $\sim 1$w & 1\\
        $\gamma$ & birth/death rate of people & $\sim 0.01y^{-1}$ & 0\\
        $K_{ab} = e^{-|b-a|/d}$ & cross-immunity of strains $a$, $b$ &  $\sim 20$ mutations &\\
        $\tau_{\rm sw}$ & coalescent time scale/sweep time & $2-6y$ &\\
        $\Tmrca$ & $T_{MRCA}$ & $\sim 3-10$y & \\
        $\dTmrca$ & $T_{MRCA}$& $\sim 2-6$y & \\
        $s$ & Selection coefficient & $\sim 0.03w^{-1}$ & $[0.003,0.05]$ \\
        $\dAnti$ & antigenic effect size & $\sim 0.5$ log2 titer approx $0.1d$ & \\
        $\mut$ & mutation rate & beneficial $10^{-3}$ per week and genome & $[10^{-7},10^{-3}]$\\
        $\Itot = \sum_a I_i$ & total prevalence & $0.005$\\
        $\mI$& average prevalence & & \\
        $N_h$ & host population & human population $10^{10}$ & $[10^6, 10^{12}]$ \\
    \end{tabular}
    \caption{{\bf Relevant quantities of influenza virus and parameters in multi-strain SIR model.}}
    \label{tab:my_label}
\end{table*}

\section*{Methods}

\subsection{Clone-based simulations}
We simulate the original model on a genealogical tree in two phases: one on the deterministic SIR-type epidemics and one on the stochastic mutation introducing new strains.
In each time step $\Delta t <1$, we apply the mid-point method to advance the epidemic equations Eqs.~(\ref{eq:infected},\ref{eq:susceptible},\ref{eq:susceptible2}). 
We then generate a random number uniformly sampled between zero and one for each surviving strain with $N_hI_a\geq 1$. If the random number is smaller than $mN_hI_a\Delta t$ for strain $a$, we append a new strain $b$ as a descendent to $a$. 
The susceptibility to strain $b$ is related to susceptibility to strain $a$ via $S_b=(S_a)^{e^{-1/d}}$.
In most of the simulations, the transmissibility of different strains is held constant $\beta$. Otherwise we allow for a strain specified transmissibility that is drawn from its parent $\beta_b=\beta_a-\delta\beta$ with $\delta\beta>0$ for the deleterious effect of antigenic mutations and $\beta_b=\beta_{\rm max}$ if the mutation is compensatory.
The new strain grows deterministically only if $\beta_bS_b>1$. 

This simplified model contains six relevant parameters: transmissibility $\beta$, recovery rate $\nu$, mutation rate of the virus $m$, birth/death rate of the hosts $\gamma$, the effective cross-immunity range $d$, and the effective size of the hosts $N_h$, whose empirical ranges are summarized in the Table.~\ref{tab:my_label}.
For flu and other asexual systems in RQS, $\beta\gtrsim\nu\gg m,\gamma$, $d\gg1$, and $N_h\gg1$.

\subsection{Fitness class-based simulations}
The stability of the RQS and the extinction dynamics is fully captured by the traveling wave equations (\ref{eq:TW}).
We simulate the traveling wave by gridding the fitness space $x$ into bins of step size $s$ around zero. The infections of different strains correspond to a natural number in each bin $x_i$.
At each time step, the population in each bin $I_i$ updates to a number sampled from the Poisson distribution with parameter $\lambda_i=I_i(1+(x_i-\bar{x})\Delta t)$ determined by mean fitness $x_i$ and a dynamic mean fitness $\bar{x}$, which increases by $\Delta tI_{tot}/N_h$, where $I_{tot}$ is the total population summed over all bins and $N_h$ is the parameter giving the host population. When $\bar{x}$ becomes larger than one bin size $s$, we shift the all populations to left by one bin and reset $\bar{x}$ to $0$, a trick to keep only a finite number of bins in the simulation.
At the same time, antigenic mutation is represented by moving a the mutated fraction in each bin to the adjacent bin on the right. The faction is determined by a random number drawn from the Poisson distribution with the mean $mI_i\Delta t$.
The typical ranges of the three parameters $s$, $m$, and $N_h$ are according to the parameters in the genealogical simulation, as documented also in Table.~\ref{tab:my_label}.

\subsection{Stochastic differential-delay simulation}
To simulate the differential delay equations Eqs.~(\ref{eq:mean},\ref{eq:nose},\ref{eq:sigma_delay}), we discretize time in increments of $\Delta t= \tau_{\rm sw}/k$ and update the dynamical variables $\chi_i=x_n(t_i)$ and $\eta_i=I_{tot}(t_i)$ via the simple Euler scheme:
\begin{eqnarray}
\chi_{i+1}&=&\chi_i+\Delta t (\chi_i-\eta_i)+\frac{\chi_i}{qs}\sqrt{\Delta t}\xi_i;\\
\eta_{i+1}&=&\bar{I}\exp\left(\tau_{\rm sw}\chi_{i-k}- \frac{\tau_{\rm sw}^2}{k} \sum_{j=0}^{k} j \eta_{i-j}\right),
\label{eq:delayapprox}
\end{eqnarray}
where  $\xi_i$ is a Gaussian random variable  with zero mean and unit variance. Mean prevalence, $\bar{I}$, enters as  the control parameter (which defines the time average of $\eta_i$).

\subsection{Influenza phylogenies}
Influenza virus HA sequences for the subtypes A/H3N2, A/H1N1, A/H1N1pdm, as well as influenza B lineages Victoria and Yamagata were downloaded from \url{fludb.org}. 

We aligned HA sequences using mafft \citep{katoh_mafft:_2002} and reconstructed phylogenies with IQ-Tree \citep{nguyen_iq-tree:_2015}.
Phylogenies were further processed and time-scaled with the augur \citep{hadfield_nextstrain:_2017} and TreeTime \citep{sagulenko_treetime:_2018}.

\section*{Appendix}
\setcounter{subsection}{0}
\subsection{Differential-delay approximation of RQS dynamics}
Here we derive the differential delay system of equations that relate the behavior of the pioneer strains with the bulk of the population. Let us consider the generating function associated with the virus fitness distribution at time $t$:
\begin{equation}
G(\lambda , t)= \sum_i I_i (t) e^{\lambda x_i (t)}
\end{equation}
where $x_i (t)=x_n(t_i)-\int_{t_i}^t dt' I_t(t')$ is the current fitness of the pioneer strain that first appeared at time $t_i$ and $I_i(t)$ is the fraction of the hosts infected by it:
\begin{equation}
I_i(t)= N^{-1}e^{ \int_{t_i}^t dt' x_t(t')}=N^{-1}e^{x_n(t_i)(t-t_i)- \int_0^{t-t_i} dt' t' \Itot(t-t')}
\end{equation}

We next take a coarse grained view of pioneer strain establishment replacing the sum in Eq. (13) by an integral over initial times $t_i \rightarrow t-\tau$
\begin{equation}
G(\lambda , t)=  \int_0^{\infty} \frac{d\tau N^{-1}}{\tau_a (t-\tau)} e^{(\tau+\lambda)x_n(t-\tau)- \int_0^{\tau} dt' (t'+ \lambda) \Itot(t-t')}
\end{equation}
Let us evaluate the integral in the saddle approximation which is dominated by $\tau=\tau^*$ corresponding to the maximum in the exponential
\begin{equation}
\tau^*+\lambda=\frac{x_n(t-\tau^*)}{x'_n(t-\tau^*)+ \Itot(t-\tau^*)} \approx \tau_{\rm sw}
\end{equation}
where we have used the deterministic limit of Eq (9). To simplify presentation we shall ignore the time dependence of $\tau_{\rm sw} =s^{-1} \log (x_n /m)$  replacing $x_n(t-\tau^*)$ in the logarithm by the time average $\bar{x_n}$.

In the saddle approximation we then have
\begin{equation}
\log NG(\lambda , t) \approx  x_n(t-\tau_{\rm sw}+\lambda)\tau_{\rm sw}- \int_0^{\tau_{\rm sw}-\lambda} dt' (t'+\lambda) \Itot(t-t')
\label{eq:logG}
\end{equation}
(where for simplicity we have omitted the logarithmic corrections). Note that by definition $G(0 , t)=\Itot(t)$.

We can now estimate fitness mean
\begin{multline}
\bar{x} (t)=\frac{d}{d \lambda } \log G(\lambda , t)|_{\lambda=0} \\
=\tau_{\rm sw}[x'_n(t-\tau_{\rm sw})+\Itot(t-\tau_{\rm sw})]- \int_0^{\tau_{\rm sw}} dt'  \Itot(t-t')\\
=x_n(t-\tau_{\rm sw})- \int_0^{\tau_{\rm sw}} dt'  I_t(t-t')
\end{multline}
and variance
\begin{multline}
\sigma^2 (t)=\frac{d^2}{d \lambda^2} \log G(\lambda , t)|_{\lambda=0}\\
=\tau_{\rm sw}[x''_n(t-\tau_{\rm sw})+\Itot'(t-\tau_{\rm sw})]+ \Itot(t-\tau_{\rm sw})
\label{eq:sigma}
\end{multline}
\EQ{sigma} involves the second derivative $x''_n$ and expect fluctuations in the establishment of new lineages (which contribute to $x'_n$) to be quite important. Yet we can get useful insight by continuing to
use the deterministic approximation to $x_n$ dynamics, in which case we arrive at simple delay relation between the variance and $x_n$
\begin{equation}
\sigma^2 (t)=\tau_{\rm sw}^{-1}x_n(t-\tau_{\rm sw})
\label{eq:sigmax}
\end{equation}
which is consistent with the variance calculated for the case of the steady TW and also satisfies the generalized  Fisher theorem
\begin{multline}
\frac{d}{dt}{\bar x} =x'_n(t-\tau_{\rm sw})+  I_t(t-\tau_{\rm sw})-I_t(t) \\
 =  \sigma^2 (t)-I_t (t)
\end{multline}

Combining Eqs.~(9,10,21) we arrive at the deterministic dynamical system approximating coupled ``ecological" SIR dynamics with the evolutionary dynamics of antigenic innovation due to the pioneer strains.
\begin{equation}
\frac{d^2}{dt^2} \log I(t) = \tau_{\rm sw}^{-1} x_n(t-\tau_{\rm sw}) - \Itot(t)
\end{equation}
\be
\frac{d}{dt} x_n(t) = \tau_{\rm sw}^{-1}x_n(t) - \Itot(t)
\label{eq:nose_eq}
\ee

This system admits a family of fixed points of the form $\tau_{\rm sw} I_{tot} = x_n ={\bar x}_n$, but as we show in the SI, the corresponding steady TW states are not always stable giving rise to limit cycle oscillations or leading to rapid extinction. Self-consistency condition relating $x_n$ and $I_{tot}$ for the steady traveling wave is readily generalized to limit cycle states. Integrating the differential-delay system over one cycle yields $\langle x_n \rangle = \tau_{\rm sw} \langle I \rangle$. An additional relation is provided by integrating $\log NG(0 , t)$ over the cycle:
\begin{equation}
\langle \log N_h \Itot\rangle = \frac{\tau_{\rm sw}^2}{2}\langle \Itot \rangle
\label{eq:self_consist}
\end{equation}

A great deal of insight into the behavior of the (deterministic) differential delay system defined above is provided by its deterministic limit (see SI) which defines the stability ``phase diagram" shown in Fig. 3(BC) that correctly captures key aspects of the behavior observed in fully stochastic simulations.


\subsection{Speciation rate as a stochastic ``First Passage" problem.}
Speciation occurs when two most distant clades persist to the antigenic independence. This persistence problem can be formulated as a first passage problem by including the second "nose" in the TW approximation.

\begin{figure}[htbp]
\centering
\includegraphics[width=.45\columnwidth]{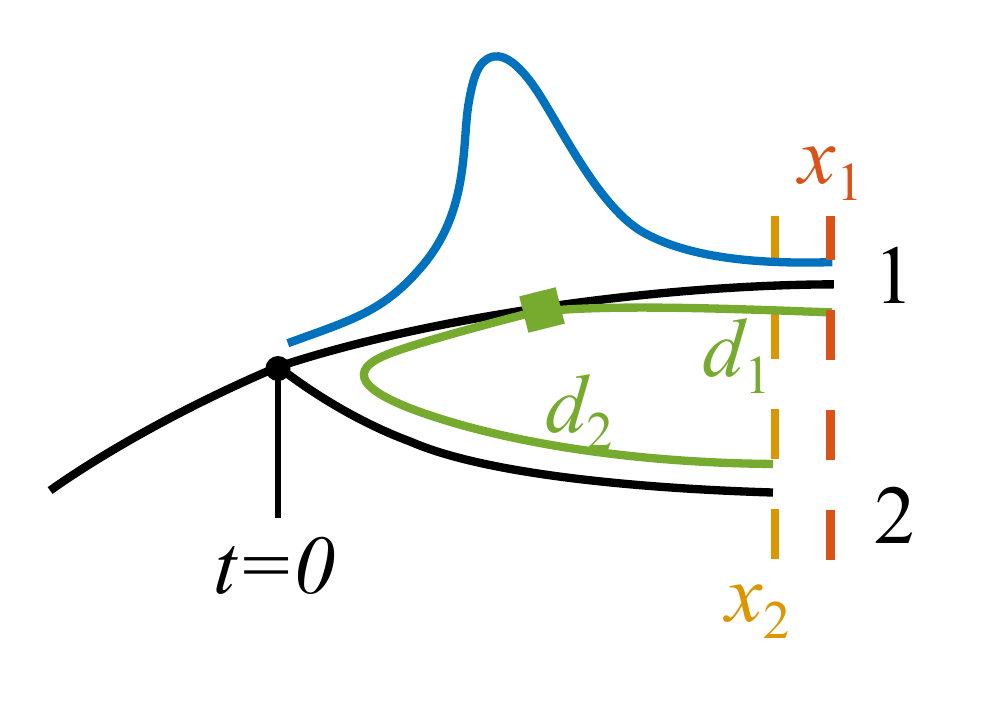}
\includegraphics[width=.53\columnwidth]{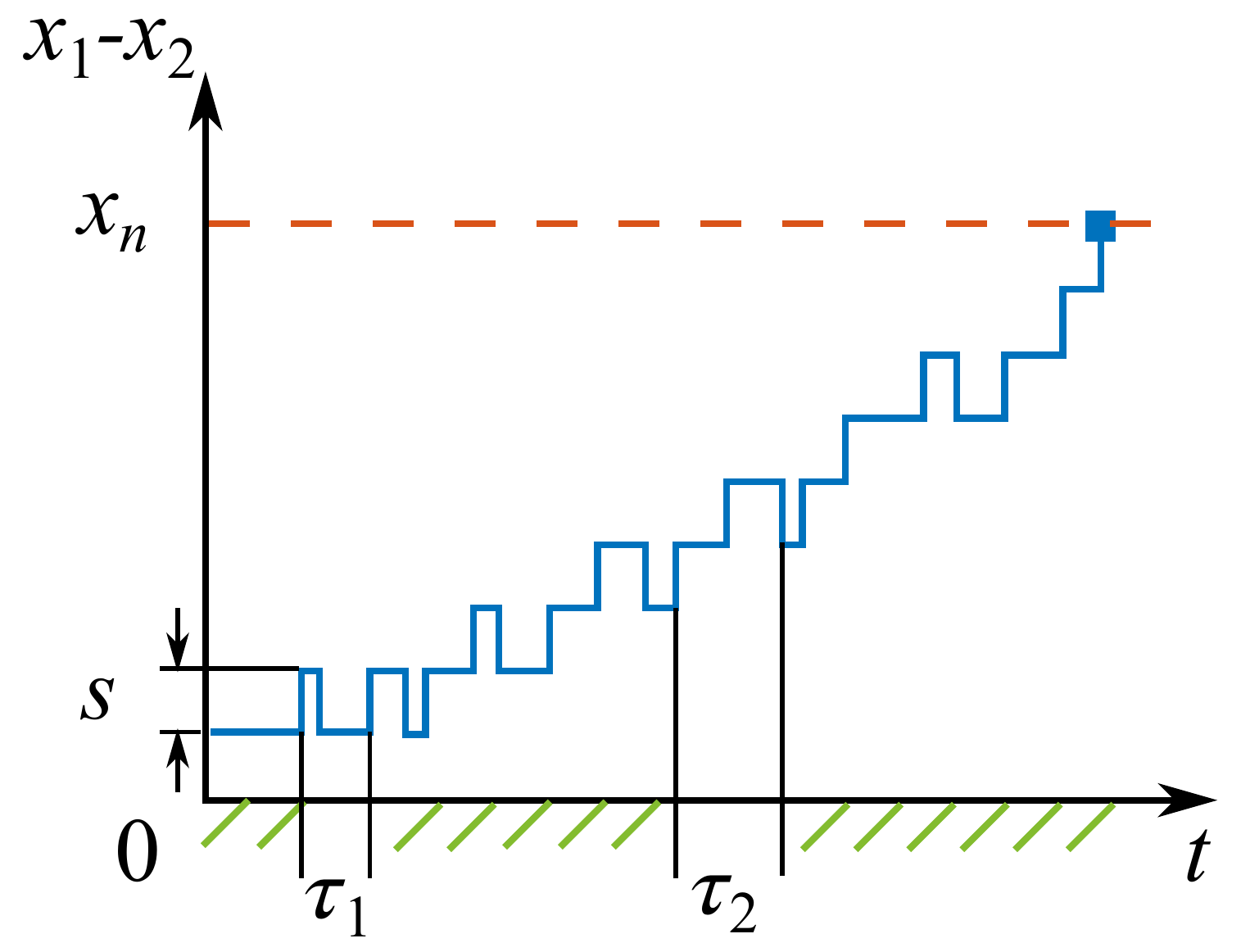}
\caption{\small{Left: Sketch of a branching event at $t=0$ with two branches 1 and 2. The fitnesses of the most fittest strains (noses) in branch 1 and 2 are $x_1$ and $x_2$. Branch 1 is the fitter one $x_1>x_2$. The antigenic distances from the cross-immune bulk to the noses of the two branches are $d_1$ and $d_2$. The Gaussian profile in fitness is illustrated in blue. Right: The fitness difference between the two branches $x_1-x_2$ is doing a biased random walk in time $t$ of step size $s$ with a reflecting boundary at $x=0$ and an absorbing boundary at $x=x_n$.
}
}\label{fig:branch}
\end{figure}

We consider the births of two pioneer strains at time $t=0$, as illustrated in Fig.~\ref{fig:branch}. The descendants of the two strains forming two branches 1 and 2 diverge in the antigenic space as they persist in time. Suppose that at time $t$, the nose of branch 1 is at fitness $x_1$, and the nose of branch 2 is at $x_2$.
Before the sweep time $t<\tau_{\rm sw}$, the cross-immunity grows mainly from the prevalent strains in the common ancestors of the two branches,
\be
\frac{d}{dt}x_i = \tau_{\rm sw}^{-1}x_i-I_{\rm tot}+s\xi_i.\qquad i =1,2
\label{eq:dx_mean}
\ee
Later when $t>\tau_{\rm sw}$, the infection bulk splits and moves on to the two branches. As the antigenic distances from the noses to the infection bulks on different branches are different, cross-immunity effects to different noses grow in different rates,
\be\ba
\frac{d}{dt}x_1&=&\tau_{\rm sw}^{-1}x_1-I_1e^{-d_{11}/d}-I_2e^{-d_{21}/d}+s\xi_1;\\
\frac{d}{dt}x_2&=&\tau_{\rm sw}^{-1}x_2-I_1e^{-d_{12}/d}-I_2e^{-d_{22}/d}+s\xi_2,
\label{eq:dx_12}
\ea\ee
where $d_{11}$ and $d_{22}$ scale roughly as $q$, the typical antigenic distance to the nose. In the limit $d_{21}\approx d_{12}\gtrsim d$, Eqs.\eqref{eq:dx_12} reduce to two independent ones of \EQ{nose} and the two branches are thus antigenically independent.
What is the probability of reaching this limit? The approach to this question rather relies on the persistence probability of two branches in the other limit when $d_{21}\approx d_{12}\lesssim d$, where $I_1+I_2\approx I_{\rm tot}$ cross-immunity growth rate is approximately the same at both noses.

In this limit, the survival probability of the less fit nose maps to a first passage problem in the random walk of relative fitness $\zeta\equiv (x_1-x_2)/x_n$. As illustrated in \FIG{branch}, an establishment of nose 1 is a positive step of $\delta\zeta=s/x_n$, while an establishment of nose 2 ends up into a backward step of the same size. As the mutations arrive in characteristic times $\tau_1$ and $\tau_2$ depending on the nose fitnesses, in the continuum limit, we have
\be
\frac{d}{dt}\zeta=\tau_{\rm sw}^{-1}\zeta+\frac{s}{x_n}\xi,
\label{eq:fitdiff}
\ee
where $\xi$ is a random noise.
There are two relevant boundaries: a reflecting boundary at $\zeta=0$ where two branches switch roles in leading the fitness, and an absorbing boundary at $\zeta=1$ where the fitness of less fit nose drops below the mean fitness and becomes destined.

The system can be solved for the probability density distribution $\rho(\zeta,t)$, which obeys
\be
\partial_t\rho(\zeta,t)=-\partial_\zeta[v(\zeta)\rho(\zeta,t)]+\partial_\zeta^2[D(\zeta)\rho(\zeta,t)],
\label{eq_FPrho}
\ee
where the drift $v$ and diffusivity $D$ depend on $\zeta$,
\be
v(\zeta)=\frac{1}{\tau_{\rm sw}}\zeta;\qquad
D(\zeta)\approx\frac{1}{q\tau_{\rm sw}}.
\ee
Solving with boundary and initial conditions,
\be\ba
\partial_\zeta\rho(\zeta,t)|_{\zeta=0}&=0;\quad\rho(\zeta=1,t)=0;\\
\rho(\zeta,t=0)&=\delta(\zeta),
\ea\ee
we have
\be
\rho(\zeta,t)=\sum_{n=1}^{\infty}e^{-\lambda_nt/\tau_{\rm sw}}c_{n1}F_1(\frac{1-\lambda_n}{2},\frac{1}{2},\frac{q}{2}\zeta^2),
\label{eq_rhozeta}
\ee
where $_1F_1$ is the generalized hypergeometric function, $\lambda_n$ is the $n$th smallest values solving $_1F_1(\frac{1-\lambda}{2},\frac{1}{2},\frac{q}{2})=0$, and coefficient $c_n$ is determined by the initial condition. In long time $t$, the slowest mode dominates the dynamics. In the large $q$ limit, we have $\lambda_1=1$.
Since $_1F_1\approx {\rm const}$ for $\zeta\in(0,1)$, the persistence probability is
\be
P(T>t)\approx c e^{-t/\tau_{\rm sw}}.
\label{eq_pdsig}
\ee
The typical time interval between successive strains scales as $\tau_a=\tau_{\rm sw}/q$. So the probability of the branch depth being deeper than $a$ is then
\be
P(D>a)\approx e^{-a\tau_a/\tau_{\rm sw}}=e^{-a/q}.
\ee
Recalling that the speciation, or escape of the cross-immunity occurs when the branch depth is larger than $d$, we find the probability of a successful branching $p_1$ is proportional to $e^{-d/q}$.

In the phylogenetic tree, $t/\tau_a$ trial branchings from the backbone arrive in time $t$. The probability that none of them successfully speciate is thus
\be
P_{\rm nsp}(t)=(1-p_1)^{t/\tau_a}=e^{-t/\tau_{\rm sp}},
\ee
where the waiting time for speciation event is
\be
\tau_{\rm sp}\propto\frac{\tau_{\rm sw}}{q}e^{d/q},
\ee
as numerically verified in \FIG{speciation}.

\setcounter{equation}{0}
\setcounter{figure}{0}
\renewcommand{\theequation}{S\arabic{equation}}
\renewcommand{\thefigure}{S\arabic{figure}}

\begin{figure}[htbp]
\centering
\includegraphics[width=1.\columnwidth]{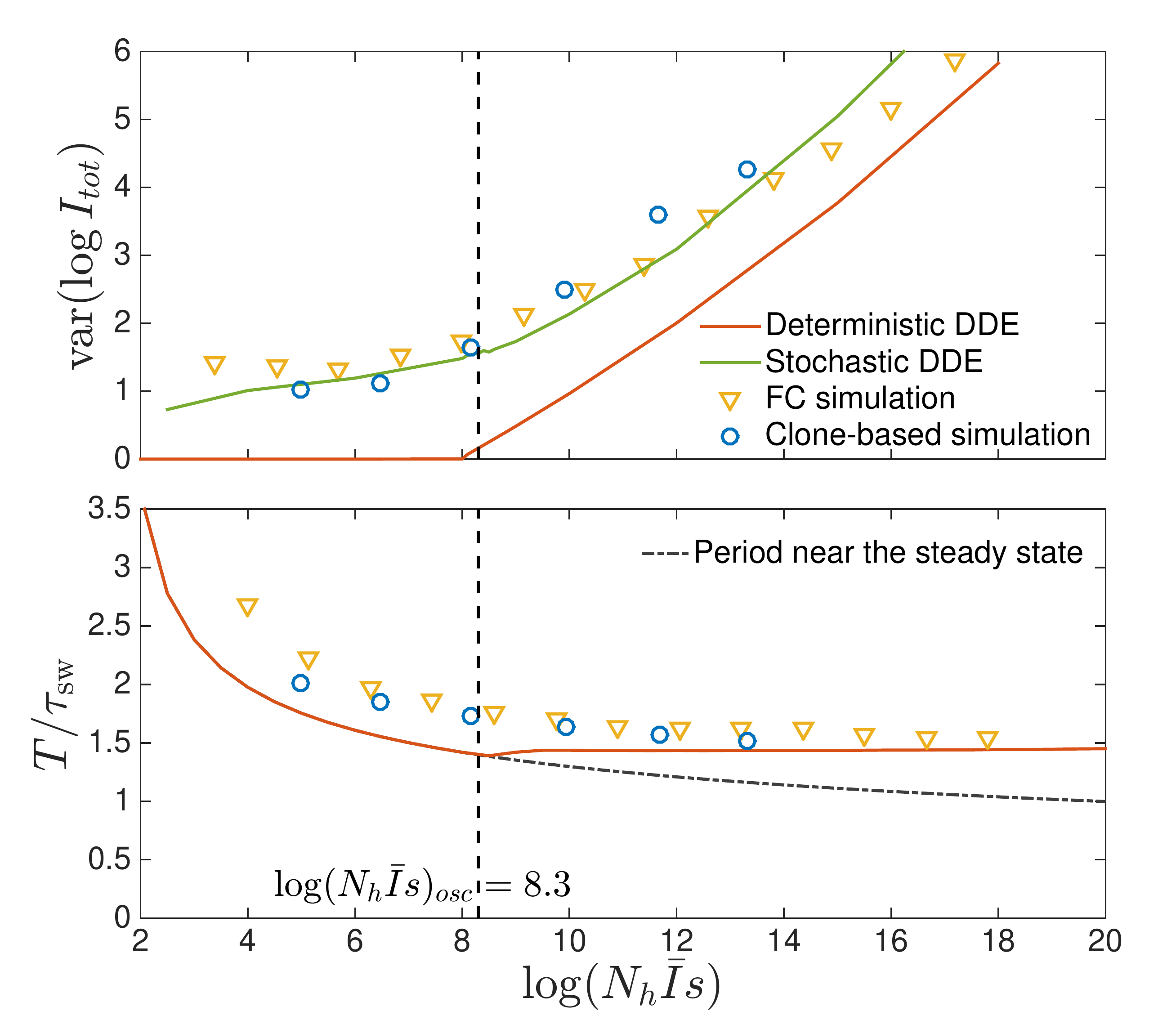}
\caption{\small{Top: Amplitude of epidemic circulations in logarithm of total prevalence. The amplitude predicted in the fitness class-based (FC) simulation is consistent with the clone-based simulation, bounded from below by the amplitude in the deterministic differential-delay approximation (DDE), which sets on at the spontaneous oscillation threshold $\log(N_h\bar{I}s)_{osc}=8.3$, indicated by the dashed line. When the noise is properly considered as in \EQ{delayapprox}, the amplitude can also be predicted from the stochastic differential-delay equations (DDE). Bottom: Period of epidemic oscillation, also bounded below by the limit cycle period of the deterministic DDE.
}
}\label{fig:variance}
\end{figure}

\section*{Supplementary Information}

\begin{figure}[htbp]
\centering
\includegraphics[width=.9\columnwidth]{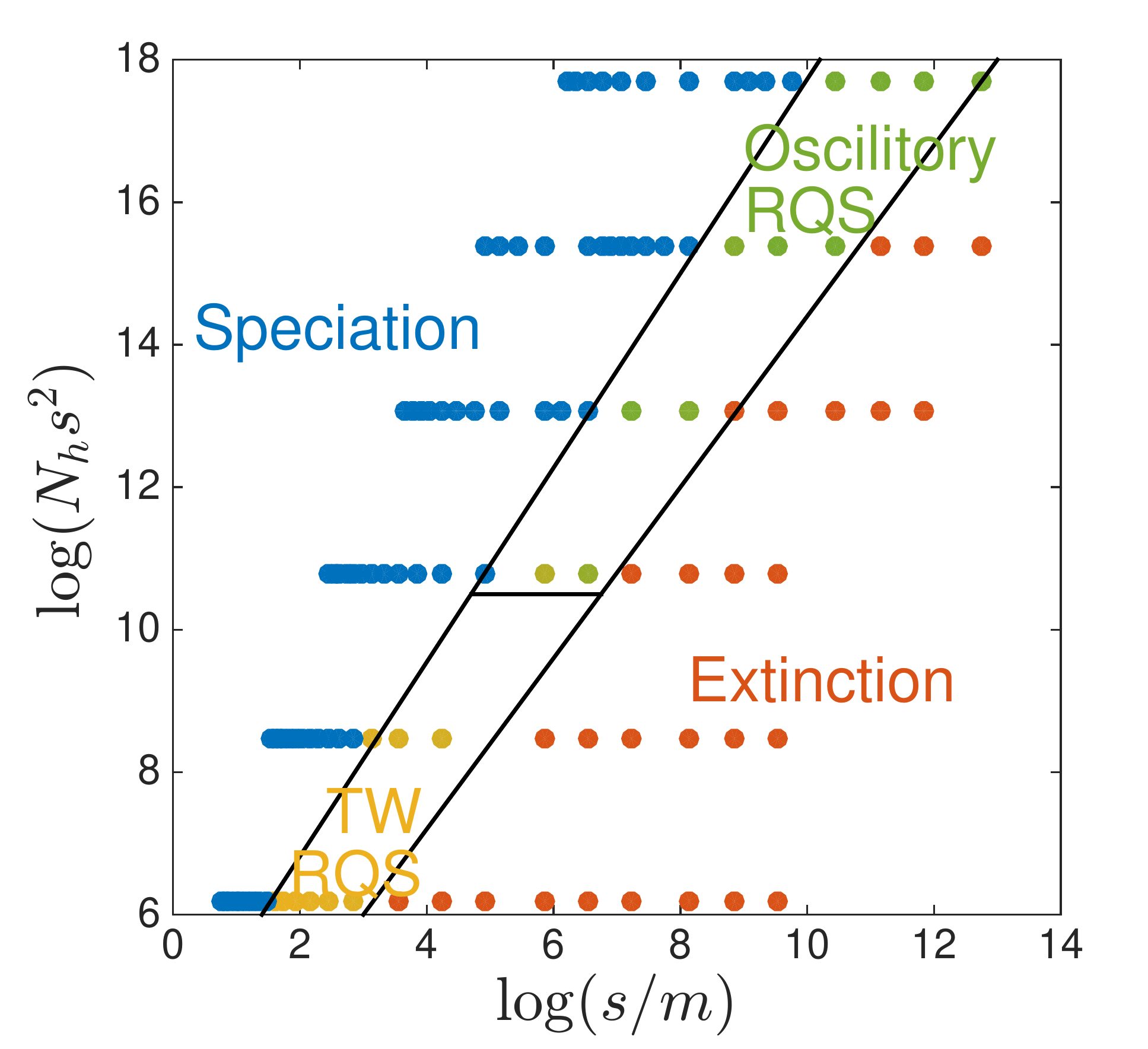}
\caption{\small{{\bf "Phase diagram" in the clone-based simulation.} The data points in corresponding phases are determined by following criteria.  The extinction phase (red): $\tau_{\rm ext}<10\tau_{\rm sw}$; The speciation phase (blue): $\tau_{\rm sp}<\tau_{\rm ext}$; The transient RQS regime (yellow and green): otherwise. Color in the transient phase labels the amplitude of prevalence ${\rm var}(\log I_{tot})$, increasing from yellow to green.
}
}\label{fig:phasedata}
\end{figure}

\subsection*{Stability analysis of the differential-delay approximation}
It is natural to measure time in the units of $\tau_{\rm sw}$: $t=\tau_{\rm sw}\zeta$, $x_n=\tau_{\rm sw}^{-1}\chi$ and define $u=\tau_{\rm sw}^2I$ and
\begin{equation}
\frac{d^2}{d\zeta^2} u(\zeta) = \chi(\zeta-1) - e^{u(\zeta)}
\end{equation}
\begin{equation}
\frac{d}{d\zeta} \chi(\zeta) = \chi(\zeta) - e^{u(\zeta)}
\end{equation}

This system has a one parameter family of fixed points $\chi=\bar{\chi}, \ u=\log \bar{\chi}$. To analyze fixed point stability we linearize and Laplace transform, yielding
\begin{equation}
z^2 \delta {\hat u}(z) = e^{-z}\delta {\hat \chi}(z) -\bar{\chi}\delta {\hat u}(z)+z \delta u(0)+\delta u'(0)
\end{equation}
\begin{equation}
z \delta {\hat \chi}(z) = \delta {\hat \chi}(za) -\bar{\chi}\delta {\hat u}(z)+z \delta \chi(0)
\end{equation}
Stability is governed by the poles of the Laplace transformed response to the initial perturbation $\delta u(0), \ \delta u'(0), \ \delta \chi(0)$ and these poles are at the complex $z$ that solve:
\begin{equation}
z  = 1+\bar{\chi}(1-z-e^{-z})/z^2
\end{equation}
Fixed point - and hence steady RQS - stability requires $\Re (z) <0$ which is found for $2<\bar{\chi}<\bar{\chi}_c $. For $\bar{\chi}>2.845$ one finds $\Im (z) \ne 0$ corresponding to the onset of oscillatory relaxation which turns into a limit cycle for $\bar{\chi}>\bar{\chi}_c \approx 16.6$. The period of the limit cycle is well approximated by $\Im (z)$, as the dashed line shown in the bottom panel of \FIG{variance}.

\subsection*{Stochastic form of the differential-delay approximation}
A sensible stochastic generalization is obtained by
the stochastic approximation for the ``nose'' dynamics (\EQ{nose})
\begin{equation}
\frac{d}{dt} x_n = \tau_{\rm sw}^{-1}x_n - I_t(t) + s \xi (t),
\end{equation}
combined with \EQ{logG} at $\lambda =0$
\begin{equation}
\log I(t) = \tau_{\rm sw}x_n(t-\tau_{\rm sw}) -\int_0^{\tau_{\rm sw}}dt't'I_t(t-t').
\end{equation}
Note that in this derivation we have avoided the need for explicitly approximating $\sigma^2$! (We have also neglected the effect of fluctuations arising from the logarithmic correction term effectively replacing it by its average value.)

\subsection*{Effect of mutations in infectivity }
Suppose an antigenic advance mutation has deleterious effect on infectivity reducing the latter by $\delta\beta_d$ on average. This would effectively reduce the fitness gain of antigenic innovation from $s$ to $s_d(\beta)=s(\beta)-\Delta_d $, with $\Delta_d=\beta^{-1}\delta\beta_d$. In addition let us assume that there also are compensatory mutations which restore maximal infectivity $\beta_{\rm max}$. These compensatory mutation thus have a beneficial effect on fitness  $\Delta_b(\beta)=\beta^{-1}\beta_{\rm max}-1$. We assume that these  mutations occur with rate  $ m_{\beta b}$.
In a dynamic balance state the rate of fixation of compensatory mutations would exactly balance the deleterious mutation effect on $\beta$ so that $\tau_{b}^{-1} \Delta_d =\tau_a^{-1} \Delta_d$ with the fixation rate controlled by the fitness of the leading strain via $\tau_{b}^{-1}=x_n/\log(\frac{x_n}{m_{\beta b}})$.
This dynamic balance is achieved at a certain value of $\beta_* <\beta_{\rm max}$, specifically $\beta_{\rm max}-\beta_*=\delta\beta_d \tau_b\tau_a^{-1}$ or
$\beta_*=\beta_{\rm max}-\delta\beta_d r$ where $r=\log(\frac{x_n}{m}) / \log(\frac{x_n}{ m_{\beta b}})$.

The fitness of the nose of the distribution obeys
\begin{equation}
\frac{d x_n}{ dt}=s_d(\beta) \tau_{a}^{-1}+\Delta_b \tau_{b}^{-1}-I
\end{equation}
where the 1st term on the RHS is rate of nose advancement due to antigenetic mutations $\tau_{a}^{-1}=x_n/\log(\frac{x_n}{ m})$ as before, but with reduced fitness gain $s_d(\beta)$. The 2nd term describes the contribution of compensatory mutations. However in the dynamic equilibrium (at $\beta_*$) compensatory mutations exactly cancel the contribution the deleterious mutation contribution to $s$ so that for the steady state we recover
\begin{equation}
\Itot=s(\beta_*) \tau_{ag}^{-1}=\frac{s (\beta_*) x_n }{ \log \frac{x_n }{ m}}
\end{equation}
as we had for the TW driven by antigenic advancement only. The only effect is the reduction of $s$ from $s(\beta_{\rm max})$ to $s(\beta_*)=d^{-1}\log\beta_*$.

The sweep time, $\tau_{\rm sw}$, upon which the fitness of the former pioneer strain comes down to the mean fitness) and the nose fitness, $x_n$,  retain the TW form
\begin{equation}
\tau_{\rm sw}=\frac{x_n}{ \Itot}=\frac{s(\beta_*) }{ \log(\frac{x_n }{ m})}
\end{equation}
Following TW approximation to estimate infection prevalence $\sqrt{\Itot} \sim N^{-1}\exp(x_n\tau_{\rm sw}/2)$ as before one finds
\begin{equation}
x_n=2\tau_{\rm sw}^{-1}\log CN_h=2  s(\beta_*)\frac{\log [ N_hs^2 /\log(\frac{x_n}{ m})]}{ \log(\frac{x_n}{m})}
\end{equation}

The total fitness variance of the population contains a contribution, from antigenic mutations and the mutations in infectivity:
\begin{equation}
\sigma^2=\tau_{\rm sw}(s_d^2 \tau_{a}^{-1}+\Delta_b^2 \tau_{b}^{-1})=x_n [s-2\Delta_d +\frac{\Delta_d^2}{ s} +\frac{\Delta_d\Delta_b}{ s} ]
\end{equation}
but under conditions of $\Delta_d, \ \Delta_b \ll s(\beta_*)$ total variance would also be decreasing.

Most relevant for our analysis however is not the typical, but the maximal antigenic distance within the viral population:

\begin{equation}
q_{ag}=\tau_{\rm sw}\tau_{ag}^{-1}=\frac{ x_n }{ s(\beta_*)}=2\frac{\log N_hs^2(\beta_*)c }{ \log \frac{x_n }{ m}}
\end{equation}
which is basically unchanged in the presence of infectivity mutations except for the expected reduction in the magnitude of $s^2$ factor inside the logarithm. Therefore, speciation rate would be reduced, but rather weakly, via a contribution subleading in $o(\log N_h)$

\bibliography{sir}
\end{document}

%% file: commands.tex
\newcommand{\EQ}[1]{Eq.~(\ref{eq:#1})}

\newcommand{\FIG}[1]{Fig.~\ref{fig:#1}}

\newcommand{\be}{\begin{equation}}
\newcommand{\ee}{\end{equation}}
\newcommand{\ba}{\begin{aligned}}
\newcommand{\ea}{\end{aligned}}
\newcommand{\bsp}{\begin{split}}
\newcommand{\esp}{\end{split}}
\newcommand{\ddt}{\frac{d}{dt}}
\newcommand{\bmm}{\begin{multline}}
\newcommand{\emm}{\end{multline}}

\newcommand{\dAnti}{\Delta}
\newcommand{\CI}{d}
\newcommand{\Tmrca}{T}
\newcommand{\dTmrca}{\delta T}
\newcommand{\Itot}{I_{tot}}
\newcommand{\mI}{\bar{I}}

\newcommand{\mut}{m}